# Reproducible Reservoir Computing with Thermally Driven Superparamagnets: Controlling Temperature Sensitivity

Zf. Chen[1], A. Welbourne[1], M. O. A. Ellis[2], D.A. Allwood[1], E. Vasilaki[2], and T. J. Hayward[1]

[1]School of Chemical, Materials and Biological Engineering, University of Sheffield, UK
[2]Department of Computer Science, University of Sheffield, UK

## Abstract:

Unconventional computing systems must demonstrate robust performance under real-world environmental conditions to enable practical deployments. We have recently proposed superparamagnetic nanodot ensembles driven by strain-induced magnetoelectric coupling as exciting candidates for use as ultra-low energy consumption reservoir computing substrates. However, because their dynamics are governed by thermal activation effects, these systems are intrinsically sensitive to ambient temperature fluctuations, leading to degraded task performance when operated outside the temperature range used during training. In this paper we simulate how temperature variations affect the magnetisation dynamics of such superparamagnetic ensembles, and quantify how this affects task performance. We then show how heterogeneous nanodot patterns that incorporate different sizes of nanodots with different characteristic timescales for thermal activation mitigate this problem. Benchmark results on the NARMA-10 task show that introducing optimised heterogeneity stabilises performance of the reservoirs across a wide range of ambient temperatures (5-35°C), with little loss of ultimate performance. We also characterise the trade-off between performance and temperature stability and show that it can be tuned via reservoir hyperparameters. Our study demonstrates a key step in making these novel devices suitable for real-world deployment.

## Introduction:

Reservoir computing (RC) is a computational paradigm particularly well suited for time-series tasks in edge computing contexts where energy footprints are limited.[1] In algorithmic form RC uses a recurrent artificial neural network with fixed internal connectivity as a dynamical "reservoir" that transform signals into a higher dimensional space where a linear classification layer can be trained to map signals to a desired output. RC has been demonstrated to be applicable to a diverse array of artificial intelligence (AI) applications, including speech recognition[2], handwritten digit image recognition[3], and digital twins[4].

To reduce the energy consumption of RC researchers have been exploring a neuromorphic variant: Physical Reservoir Computing (PRC). Here, instead of algorithmically simulating a reservoir, one can exploit the natural dynamics of a physical system to perform to provide a similar transformation.[5] This shift is driven by an important insight: the inherent 'black-box' nature of a reservoir implies that any dynamical system offering the right dynamical features can serve as a reservoir.[6] To act as a reservoir a material should: (a) exhibit nonlinear responses to external stimuli; (b) possess a rich state space that allows input to be projected into higher dimensions and (c) demonstrate fading memory, where the effects of past stimuli are gradually diminish as new inputs are provided. [7]Studies have demonstrated physical reservoir computing (PRC) in a wide array of physical substrates including e.g. photonic, robotic, and spintronic systems. [8–13]

Given RCs applicability to simple tasks at the edge, finding systems that are inherently energy efficient is paramount. In a previous study, we addressed this by proposing a novel reservoir based on a superparamagnetic nanodot ensemble driven by strain-induced magnetoelectric coupling.[14] Our numerical simulations demonstrated the theoretical feasibility of such a physical reservoir, and showed that it could exhibit both competitive computing performance and ultra-low energy consumption.[14] However, despite the promise of this approach, the reservoir's key strength, i.e. its low energy, thermally-driven dynamics also create a key weakness: the system's dynamics will be highly sensitive to ambient temperature changes creating inference errors when deployed in a real-world environment.

In this work, we will address this limitation by demonstrating how heterogeneous reservoirs that contain nanodots with a range of sizes can be used to stabilise computational performance under ambient temperature variations. We begin by exploring how the superparamagnetic ensemble's dynamical behaviours change with temperature and perform a systematic evaluation of task performance when temperature varies from that under which the reservoir was trained. We then use a Bayesian optimization methodology to develop a heterogeneous nanodot patterns that are optimised to provide both strong task performance and excellent average NMRSE across an appropriate temperature range (5°C-35°C). Finally, we show how tuning the feedback strength of the reservoir allows us to adapt the reservoir to trade off minimum NMRSE against the temperature stability, allowing the reservoir to be tuned for the environment under which it is deployed.

## Methodology:

In Fig.1(a), we show a schematic diagram of the superparamagnetic nanodot ensemble based reservoir, as presented in our previous publication[14].

The proposed reservoir consists of a large ensemble of circular CoFeB nanodots (here we assume >1000 dots, spaced by >2x their diameter to minimize dipolar coupling) on top of a piezoelectric substrate. The piezoelectric substrate (PMN-PT) undergoes either compressive or tensile deformation due to the inverse piezoelectric effect in response to changes in applied voltage.[15] The deformation-induced strain is elastically transferred to the CoFeB superparamagnetic nanodots, modifying the height and symmetry of the energy barrier between bistable magnetisation states within the nanodots, which have uniaxial anisotropy.[16] This change modifies the rates of thermally activated magnetisation switching between the nanodot's bistable states, as governed by the Néel–Arrhenius law.[17] For an extended ensemble, these individual stochastic switching manifest as a predictable collective change in net magnetisation that exhibits key features that align with the fundamental requirements of a physical reservoir, including reproducibility, non-linearity, and fading memory.[7] More details have been provided in our previous work.[14]

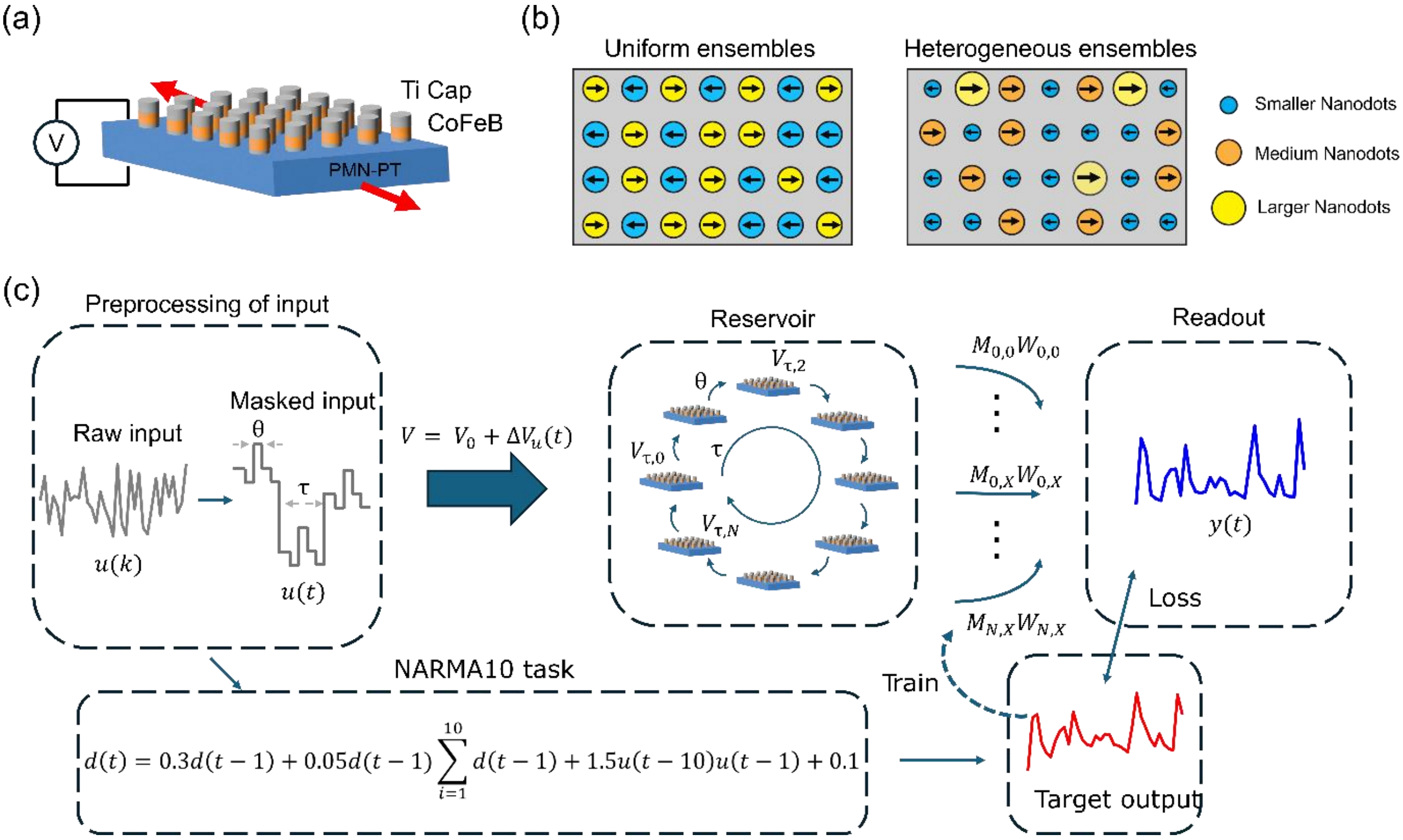

Fig.1. (a) Schematic of a superparamagnetic nanodot ensemble consisting of a PMN-PT substrate onto which CoFeB/Ti dots are patterned. (b) Schematic diagrams illustrating uniform and heterogeneous nanodot ensembles. (c) Schematic representation of the time-multiplexed reservoir computing process for NARMA-10 task. Raw inputs are masked by multiplying with a binary matrix. The masked inputs are transferred into the reservoir along with feedback, where the feedback is the product of the output from each virtual node one time step previously and the feedback strength $\gamma$. The output layer is trained via linear regression to minimise loss against the target output of the NARMA-10 task.

In the following we describe the model used to simulate the thermally driven magnetisation dynamics of the nanodot arrays. Here, the nanodots are assumed to be composed of CoFeB with saturation magnetisation $M_s = 7.2 \times 10^5 Am^{-1}$, exchange stiffness $A_{ex}$=10$pJm^{-1}$, and a growth-induced uniaxial anisotropy $K = 1.2 \times 10^3 Jm^{-3}$[18,19].

For strain-mediated, voltage-controlled CoFeB nanodots, the energy under an applied field can be described by the Stoner–Wohlfarth model:

$$E(\theta) = K_1 V sin^2(\theta) + K_\sigma V sin^2(\theta - \phi) - \mu_0 M_s H V cos(\theta - \rho) \qquad \text{(equation 1)}$$

where $K_1$ is the intrinsic uniaxial anisotropy constant, $V$ is the volume of the nanodot, $\theta$ is the angle of magnetisation with respect to the uniaxial anisotropy axis, $K_\sigma = \frac{3}{2}\lambda_s E\varepsilon$ represents the strain-induced anisotropy, where $\lambda_s$ is the saturation magnetostriction coefficient, $E$ is the Young's modulus of CoFeB and $\varepsilon$ is the voltage-induced strain of the substrate. $\mu_0$ is the vacuum magnetic permeability, $M_s$ is the saturation magnetisation, $H$ is the strength of the external magnetic field and $\rho$ is the angle of the magnetic field to the anisotropy axis.

Applications of voltage alter the angular-dependent energy landscape of CoFeB particles, making the energies of the bistable ground states and the directional energy barriers $\Delta E^{ij}$ between them unequal. In the presence of thermal fluctuations, each nanodot's magnetisation stochastically switches between two states on a characteristic timescale, $\tau = \frac{1}{\omega}$. The transition rate $\omega$ from state $i$ to state $j$ follows the Néel–Arrhenius law, as shown below:

$$\omega_{ij} = f_0{}^{ij} e^{-\frac{\Delta E^{ij}}{k_B T}} \qquad \text{(equation 2)}$$

where $f_0{}^{ij}$ is the attempt frequency(typically $10^{-9}$s[20]); $\Delta E^{ij}$ is the energy barrier from state $i$ to $j$; $k_B$ is the Boltzmann constant and $T$ is temperature.

The time evolution of the net magnetisation of the ensemble is then given by:

$$m(t) = m_0 e^{-t/\tau} + m_{eq}(1 - e^{-t/\tau}) \qquad \text{(equation 3)}$$
$$m_{eq} = (\omega_{21} cos\theta_1 + \omega_{12} cos\theta_2)/\omega \qquad \text{(equation 4)}$$

where $m_0$ is the initial magnetisation of the system ($t = 0$), $m_{eq}$is the magnetisation at equilibrium state, $\theta_1$ and $\theta_2$ characterise the angles corresponding to the two ground states.

The key hypothesis of this paper is that the dynamics of the nanodot ensembles will exhibit reduced temperature sensitivity if they incorporate heterogeneous distributions of dot sizes that introduces multiple characteristic timescales. To simulate the magnetisation $m_{hetero}$of heterogeneous nanodot ensembles, we employed a weighted-averaging approach. We modelled multiple uniform ensembles with different nanodot diameters under identical input conditions and collected their respective magnetisation $m_k$. The $m_{hetero}$ was then calculated as the weighted sum of these individual ensemble magnetisations, using the number factions $w_k$ corresponding to each nanodot size:

$$m_{het} = \sum_k w_k m_k \quad \text{(equation 5)}$$
$$\sum_k w_k = 1 \qquad \text{(equation 6)}$$

where $w_k$ denotes the number fraction of the ensemble $k$.

Fig.1(c) illustrates our reservoir computing process. This follows the approach of Appeltant [21] where a reservoir is formed from a single dynamical node via time multiplexing data into a stream of *N* virtual nodes. In the preprocessing of input, the discretised raw input signal $u(k)$ (in the case of the tasks here: a one-dimensional, time-discrete sequence) is converted into a time-sequential masked input $S_n$ via multiplication with a mask $M_{d,N}$ (a matrix of shape $d * N$, where $d$ represents the dimensionality of the input).[21] The result of this is a stream of *N* virtual node inputs per point in $u(k)$. This data is passed into the reservoir sequentially, with each value being held for time $\theta$ denoting the duration of each virtual node. Thus, $\mathrm{T} = N \times \theta$ is the duration over which virtual nodes are processing a single input from $u(k)$ .
$S_n$ is applied to the reservoir as a voltage applied to the piezoelectric substrate, and is augmented with an additive feedback term that connects virtual nodes across input samples. For a sample index $n$ and virtual node $i$, the input is described by the following equation:

$$v_{ni} = \Delta v S_{ni} + \gamma x_{(n-1)i} \qquad \text{(equation 7)}$$

where $\Delta v$ is a scaling factor to give an input voltage into the reservoir, $\gamma$ is the feedback strength, and $x_{(n-1)i}$ is the output of the reservoir from virtual node *i* for input sample $n-1$.

The output $x_{ni}$ , for the $n$th sample and $i$th virtual node, of the reservoir is measured at the end the each $\theta$ period as:

$$x_{ni} = m_x(v_{ni}) \quad \text{(equation 8)}$$

The outputs across all $i$ virtual nodes then form the reservoir state matrix for a given input sample:

$$X_n = \begin{bmatrix} x_{n0} \\ x_{n1} \\ \vdots \\ x_{nN} \end{bmatrix} \qquad \text{(equation 9)}$$

The output $Y$ is obtained by combining the state matrix $X$ with trained linear weight matrix $W$:

$$Y = XW \;\; \text{(equation 10)}$$

W is found via ridge regression[22]:

$$W = Y_{train}X^T(XX^T + \lambda I)^{-1} \;\; \text{(equation 11)}$$

Where $Y_{train}$ is target output from the training dataset, the $\lambda$ is the L2-norm regularization hyper-parameter and $I$ represents the identity matrix.

In this study, we use the well-known NARMA-10 (Normalized Auto-Regressive Moving Average)[23,24] task to evaluate the prediction performance of the reservoir. The raw input comprises 3,000 samples distributed in the range of [0,0.5], with 2,000 samples for training and 1,000 samples for testing. The output $y_n$ is generated via Eq. (12), as illustrated in Fig.1(c).

$$y_n = y_{n-1}\left(0.3 + 0.05\sum_{k=1}^{10} y_{n-k}\right) + 1.05u_{n-1}u_{n-10} + 0.1 \qquad \text{(equation 12)}$$

Performance is evaluated using the Normalized Root Mean Square Error (NRMSE):

$$NRMSE = \sqrt{\frac{1}{N_s}\frac{\sum_n^{N_s}(\widetilde{y_n} - y_n)^2}{Var(y)}} \;\; \text{(equation 13)}$$

where $N_s$reflects the sample number in the testing, $y_n$ is the target output and $\tilde{y}_n$ is the output from the reservoir.

## Results and discussion

We first illustrate how temperature affects the dynamics behaviour of superparamagnetic nanodots (Fig.2(a)). As temperature increases the rates of transitions ($\omega_{12}$and $\omega_{21}$) between bistable states increases, allowing equilibrium across a population of nanodots to be reached more quickly as the energy landscape evolves. To quantify this, we calculated value variations in the key the transition rate $\omega$, characteristic timescale $\tau$ and magnetisation $M$ dynamics, induced by temperature changes.

Fig.2(b) shows $\omega$ and $\tau$, as a function of temperature for an individual CoFeB particle as given by Eq.(2) for $f_0 = 10^{-9}s$. Here, the CoFeB particle is set as a 50nm diameter, 4nm thickness nanodot with $M_s = 7.2 \times 10^5 Am^{-1}$ and $K = 1.2 \times 10^3 Jm^{-3}$. In the tested temperature range, the transition rate ω increases (equivalently, the characteristic timescale τ decreases) exponentially with temperature. Specifically, ω increases by nearly 13 times as temperature rises from 270 K to 310 K (typical ambient temperature variation), which corresponds to a reduction of $\tau$ from about 11 μs to 0.7 μs. This indicates that both the timescales of dynamics vary strongly with temperature, even for relatively small temperature changes.

Fig.2(c) presents the simulated magnetisation evolution of a nanodot ensemble subjected to an identical series of voltage/strain inputs under three different temperature conditions (6°C, 20°C and 35.4°C). Here, input is defined as the ratio $r = K_\sigma/K$ , where $K_\sigma = \frac{3}{2}\lambda_s E \varepsilon$ is strain-induced anisotropy, $K$ is the intrinsic anisotropy. The input is chosen as $r \in [-0.4, 0.4]$. Using the material parameter ($\lambda_s$=60ppm[25], $E$=216GPa[26]), the input amplitude of $r = \pm 0.4$ corresponds to a strain of $\varepsilon \approx 2.47 \times 10^{-5}$(0.00247%). The strain is much smaller than the maximum in-plane strain (~0.175%[15]) that can be generated in PMN-PT. The magnetisation curve under 20°C is used as a reference. For the higher temperature (35.4°C), the magnetisation deviates from the reference with a maximum deviation of 30% at the end of the first input step.  Across the full sequence, we can take the NRMSE (or deviation) as a useful measure of the deviation between the curves: this is 20% as given by Eq. (13). Similarly, the magnetisations at 6°C were observed to also deviate from the reference but in the opposite direction. The maximum deviation is 26.7% and the NRMSE is 14%.

To better understand these temperature-induced deviations, we next analyse the underlying temperature-dependent magnetization dynamics of the system, taking the 20 °C trace (blue curve in Fig.2(c)) as a reference. The system's magnetisation begins to evolve from an initial state toward an equilibrium state via thermal relaxation once the first voltage (or strain) stimulus is applied. This evolution occurs because individual nanodots flip between bistable states, with an imbalance in switching probabilities due to the asymmetry of two energy barriers, which are modulated by the applied stimulus. The system does not fully equilibrate before the second stimulus is applied. Upon the application of the second stimulus, there is a very rapid change in magnetisation. This is a result of the shift in the angular locations of the two bistable minima, and the magnetisations of nanodots reorient rapidly to minimise their energy. Subsequently, the system undergoes thermal relaxation to reach a new equilibrium state again. Notably, temperature strongly affects the thermal relaxation process, determining how rapidly the system evolves toward equilibrium after each stimulus. Under warmer conditions, as shown by the orange curve in Fig.2(c), flipping over the energy barrier within nanodots becomes easier, allowing the system to be closer to obtaining equilibrium in each input step and resulting in a stronger collective magnetisation response. Conversely, under cooler conditions, the relaxation becomes slower, leading to weaker dynamics, as illustrated by the green curve in Fig.2(c).

These results highlight the critical role of temperature in determining the computing performance of the proposed nanodot reservoir: dynamics of the nanodot system exhibits a strong temperature dependence, leading to distinct magnetisation responses to identical input sequences under different thermal conditions. This discrepancy suggests that our superparamagnetic reservoir may suffer from poor computational reproducibility even under relatively small temperature variations such as one might see over e.g. a day/night cycle.

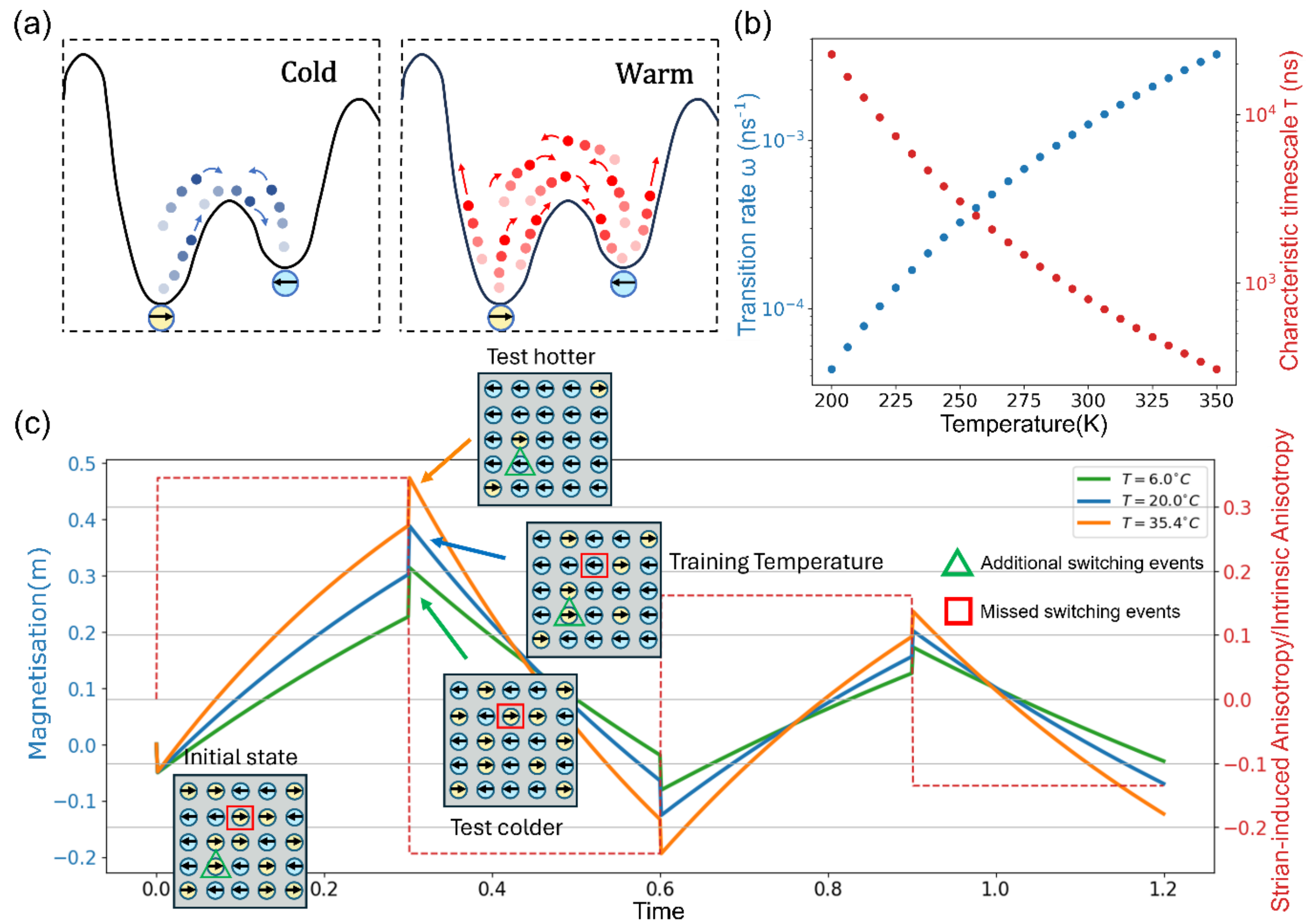

Fig.2. (a) Schematic illustrating the thermal activation behaviour of single-domain nanodots at two different temperatures. The probability of magnetisation switching is increased at higher temperatures.(b) Transition rates $\omega$ and characteristic timescales $\tau$ of nanodots as a function of temperature in a single 50 nm diameter, 4nm thickness CoFeB nanodot with a field (h=0.4) applied at 90° to the anisotropy axis, as given by Eq.(2) for $f_0 = 10^{-9}s$, (c) Dynamics of a nanodot ensembles net magnetisation over time in response to an input sequence under different temperatures (6.0°C, 20°C and 35.4°C). Input (Strain-induced anisotropy $K_\sigma$/ intrinsic anisotropy $K$) selected randomly from -0.4 to 0.4.

To quantify how the variation of reservoir dynamics with temperature affected computational performance, we have evaluated the performance of a uniform nanodot reservoir on the NARMA-10 benchmark over a temperature range of 5 - 35°C. Here we performed simulations with the uniform array that achieved the best prediction performance in our hyperparameter sweep (NRMSE = 0.49; see Supplementary Material for details). In all cases inference is performed with the weights matrix (W) that was trained 20°C, thus replicating a circumstance where a reservoir is trained for a task under one set of conditions, but deployed under another. Here, we define the relative error as the relative deviation between the NRMSE obtained at the tested temperature and that obtained at 20 °C (as expressed in Eq. (13))

$$Relative\ Error\ = (NRMSE_{test} - NRMSE_{20°\mathrm{C}})/NRMSE_{20°\mathrm{C}} \qquad \text{(equation 14)}$$

Fig.3(a)&(b) plots NRMSE and relative error as a function of inference temperature. Asymmetric changes in NRMSE accompanied temperature deviation from the training temperature: as the temperature decreases, the NRMSE increases rapidly, reaching 550 at 5°C, whereas the NRMSE also increases with the temperature increase but less significantly with NRMSE of 6.14 at 35°C. Even within a narrow temperature window (±4°C from training temp), as presented in Fig.3(b), the NRMSE still surpasses 1.1 (with relative error is over 140%). This confirms that the uniform nanodot ensemble is highly temperature-sensitive and thus cannot perform well as a reservoir unless operated under an ultra-stable thermal environment.

To provide further insight Fig.3(c) compares the target NARMA-10 output and reservoir reconstruction at three temperatures: 13.6°C (purple triangle), 20°C (deep blue circle) and 29.1°C (green diamond). At the training temperature of 20°C, the reservoir's reconstruction is highly consistent with the target. By contrast, at 13.6°C the reservoir reconstruction shows virtually no correspondence to the target, indicating that the reservoir entirely loses predictive capability. The slower speed of the magnetisation dynamics can be clearly seen in the manner in which the reconstructed trace drifts slowly around the target. At 29.1°C, the reservoir is more responsive, as would be expected at a higher temperature discussed, however it fails to correlate well with the target.

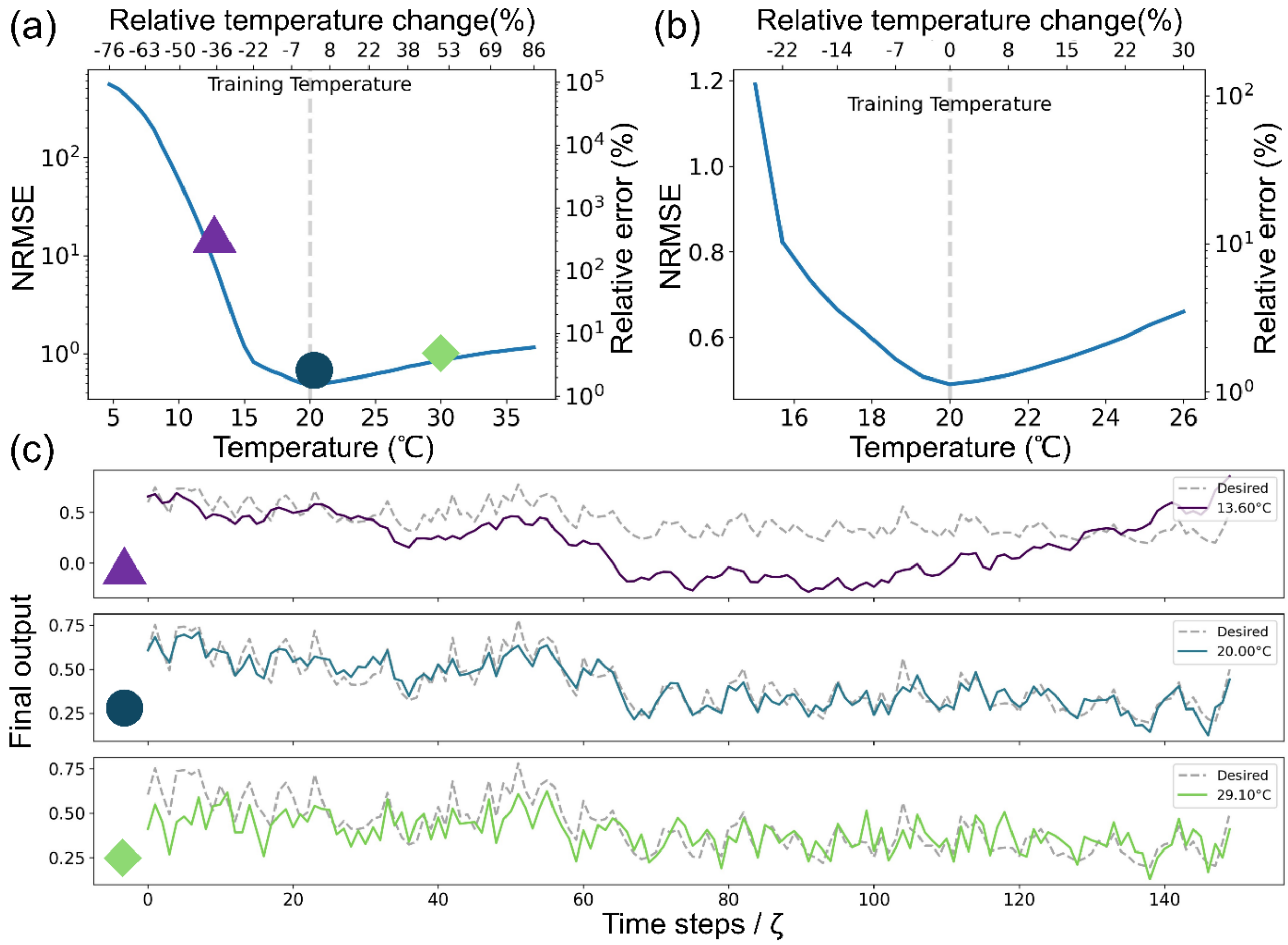

Fig.3. (a) NRMSE and relative error of the reservoir in the NARMA-10 task as a function of temperatures in the range 5 - 35°C. The reservoir exhibits optimal prediction performance at the training temperature, with performance progressively degrading as the temperature deviates from 20°C. (b) The same data plotted over the range 15-26°C. (c) Comparison of the target output and reservoir prediction output at three temperatures. Grey dashed lines indicate the desired output from the NARMA-10 task.

Clearly, uniform nanodot reservoir fails to yield reproducible outputs under thermal fluctuations. Introducing geometric heterogeneity is expected to generate a distribution of magnetic switching timescales. This diversity in dynamical response is expected to moderate temperature-induced variations of the ensemble as a whole, providing more consistent responses across temperatures. However, designing such a system presents two significant challenges: Firstly, heterogeneous reservoirs exist in much a larger phase space than homogeneous reservoirs, making the system design and optimisation more difficult. Secondly, reservoirs must simultaneously exhibit both strong peak task performance at the training temperature *and* stable performance when operating away from that temperature, necessitating Multi-Objective Optimization[27,28].

To address these challenges Multi-Objective Bayesian Optimization (MOBO)[29] is used to co-optimise the geometry design and hyperparameters of reservoirs. MOBO searches parameter space of a system to optimise the vector objective function $y_x = \{y_1(x), y_2(x), \dots, y_n(x)\}$ through a loop of sampling, evaluation and function updating. Generally, no single solution $x^*$ is optimal for all objectives simultaneously. The algorithms therefore provide a set of non-dominated solutions, known as Pareto front, in which no single objective can be improved without degrading at least one other objective.[30] We used Optuna[31], an automatic hyperparameter optimization framework, to obtain the pareto front of our problem. In this study, we define two competing objectives with respect to the computational capability of reservoirs subjected to temperature variation, namely Minimum NMRSE and the average NMRSE of performance. Minimum NMRSE characterises the optimal computational performance that the reservoir can offer under temperature variations, quantified by the minimum NRMSE observed across the tested temperature range, whereas average NMRSE describes how consistently the reservoir maintains its computational performance under thermal variations and it is evaluated as the average NRMSE for the reservoir across the temperature range under test. To facilitate comparison with the results of the uniform reservoir discussed previously, the selected temperature range was set to 5–35 °C. All input parameters and their ranges are shown in Table Ⅰ. A NSGAII sampler implemented was used to efficiently search the hyperparameter space.[31]

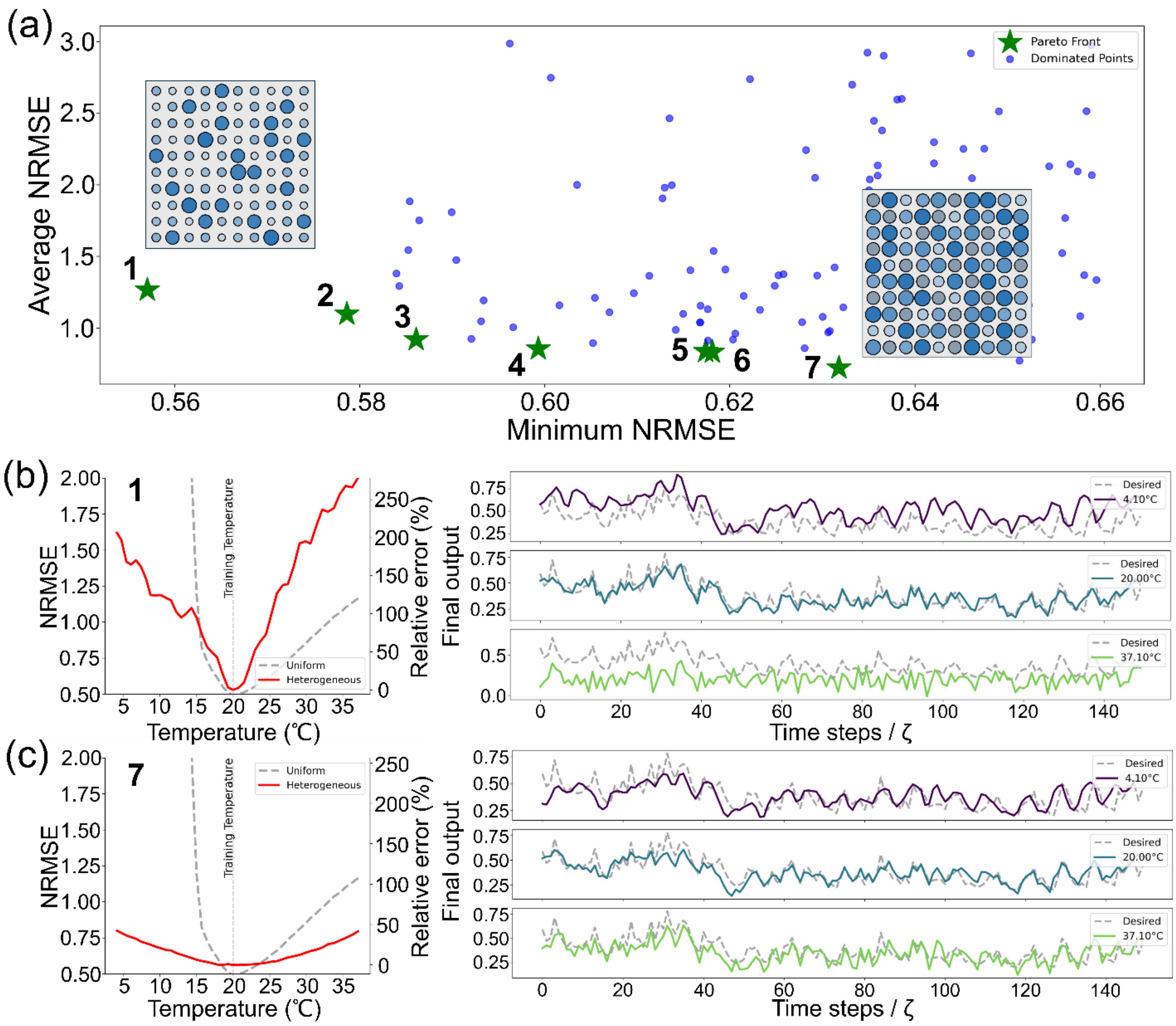

Fig.4. (a) Optimization results presented in the objective (minimum NMRSE- average NMRSE) space. Pareto front points are denoted by green stars and labelled sequentially (1,2,3...) for reference, demonstrating an approximately linear trade-off relationship between two competing objectives. Insets are illustrations of No.1 and 7 heterogeneous nanodot reservoirs for illustrating nanodot size variation marked by different colour. For simplicity, each schematic is presented as a 10*10 configuration. The geometric configurations of all reservoirs

on the Pareto front are summarized in Table Ⅱ. (b)-(c) Reservoir performance on NARMA-10 tasks across a wide-temperature range (5-35°C). The left panels depict the variations in NRMSE and relative error as functions of temperatures, while the right panels provide the comparison between the target output and reservoir prediction output. (b) corresponds to the reservoir with optimal peak (No.1) performance in (a) and (c) represents the reservoir with best average NMRSE (No.7) in (a). Parameters for reservoirs on the Pareto front were listed in Table Ⅲ.

Table Ⅰ: Parameter ranges explored by MOBO in NARMA-10 task

| Category | Parameter | Unit | Range |
|---|---|---|---|
| Geometric factors | Nanodot species | integer | 2 - 7 |
| | Diameter ratio ($d_2/d_1$) | - | 0.75 – 1.25 |
| | Proportion of species | (%) | 0 – 100 |
| Hyperparameters | Virtual node number ($N_v$) | integer | 20 - 500 |
| | Feedback strength ($\gamma$) | - | 0.05 - 2 |
| | Applied field ($h$) | - | 0.3 – 0.5 |
| | Input scaling ($\Delta(v)$) | - | 0.001 – 0.005 |
| | Input rate ($\theta/T_0$) | - | 0.2 – 0.4 |

Note. d1 and d2 represent diameters of heterogeneous nanodots and 50nm nanodots, separately. $T_0$ is defined as the system's characteristic timescale under zero-input (or strain) conditions.

Table Ⅱ Geometric Parameters for reservoirs on Pareto front

| No. | Nanodot species | Diameter ratio ($d_2/d_1$) | Proportion of species (%) |
|---|---|---|---|
| 1 | 5 | [0.81,0.86,0.87,1.14,1.18] | [20, 28, 31,18, 2] |
| 2 | 6 | [0.83, 0.84,0.96,0.98,1.16,1.2] | [20, 13, 8, 17, 19, 23] |
| 3 | 6 | [0.78, 0.86, 0.88, 1.04, 1.16, 1.2] | [14, 17, 9, 22, 26, 11] |
| 4 | 4 | [0.75, 1, 1.01, 1.16] | [1, 29, 38, 32] |
| 5 | 5 | [0.75, 0.78, 0.84, 0.85, 0.99] | [22, 11, 27, 11, 29] |

| 6 | 6 | [0.82, 0.89, 0.91, 0.98, 1.06, 1.12] | [9, 19, 6, 21, 20, 25] |
|---|---|---|---|
| 7 | 6 | [1.01, 1.11, 1.12, 1.2, 1.23, 1.24] | [16, 20, 20, 15, 10, 18] |

Note. d1 and d2 represent diameters of heterogeneous nanodots and 50nm nanodots, separately.

Table III Hyperparameters parameters for reservoirs on Pareto front

| | Hyperparameters | | | | | Objectives | |
|---|---|---|---|---|---|---|---|
| No. | Virtual node number ($N_v$) | Feedback strength ( $\gamma$) | Applied field ( $h$) | Input scaling ( $\Delta(v)$) | Input rate ($\theta/\mathrm{T}_0$) | Minimum NMRSE | Average NMRSE |
| 1 | 149 | 0.1555 | 0.4291 | 0.0018 | 0.2057 | 0.557 | 1.265 |
| 2 | 83 | 0.1555 | 0.4567 | 0.0035 | 0.3861 | 0.578 | 1.097 |
| 3 | 83 | 0.1555 | 0.4567 | 0.0035 | 0.3861 | 0.586 | 0.918 |
| 4 | 154 | 0.1350 | 0.4375 | 0.0021 | 0.2232 | 0.599 | 0.854 |
| 5 | 46 | 0.1587 | 0.4895 | 0.0021 | 0.2877 | 0.617 | 0.835 |
| 6 | 152 | 0.1555 | 0.4156 | 0.0017 | 0.3861 | 0.618 | 0.831 |
| 7 | 358 | 0.1189 | 0.4567 | 0.0021 | 0.3791 | 0.631 | 0.720 |

Fig.4(a) shows the solution set obtained after 400 trials, illustrating both the Pareto front and dozens of dominated points. For clarity, the solutions on the Pareto front are sequentially labelled and are referred to by these indices in the following discussion. The front clearly reveals a near-linear trade-off between two objectives: the best minimum NMRSE (NRMSE = 0.557) is attained at average NMRSE = 1.265, whereas the greatest average NMRSE (average NRMSE = 0.720) is achieved with a minimum NMRSE of 0.631. In contrast the equivalent number for the best homogeneous reservoir is 69.861. Collectively these findings support our hypothesis that the heterogeneous reservoir offers better average NMRSE across a wide-temperature range.

Fig.4(b) and (c) compare two heterogeneous nanodot reservoirs (No.1 and No.7). The left panels show the variations in NRMSE and relative error as functions of temperatures of the reservoir, while right panels present the reservoir output traces for the NARMA-10 task at three temperatures (4.1°C, 20°C and 37.1°C) against the target output trace. As shown in Fig.4(b), the No.1 reservoir exhibits minimum NMRSE of 0.557, comparable to that of the uniform reservoir (NRMSE = 0.49). This indicates the introduction of the geometric heterogeneity does not inherently lead to significant deterioration of the minimum NMRSE. However, this reservoir shows poor stability of performance as temperature changes; the output trace reproduces the target output accurately at 20°C, but exhibits large deviations from the desired signal at 4.1°C and 37.1°C. In contrast, reservoir No.7 (Fig.4(c)), which provides the most thermally stable performance on the Pareto front, shows minimum NRMSE of 0.631 and average NRMSE of 0.720. This is consistent with trace comparisons presented on the right panels of Fig.4(c). All three output traces show good agreement with the target output, exhibiting only minor deviations.

The above results demonstrate that the heterogeneous nanodot reservoirs are capable of being optimised to show both strong minimum NMRSE and exhibit much higher average NMRSE than homogeneous systems. However, for a real device it would be helpful to tune the trade-off between these two performance metrics situationally, which cannot be achieved by modifying geometry. Therefore, we next focus on investigating how hyperparameters influence this trade-off, establishing firstly an importance analysis of the available hyperparameter and then exploring how these could be used to enable adaptation of a specific reservoir. Here, we chose No.7 reservoir with the lowest average NRMSE in Fig.4 as the subject.

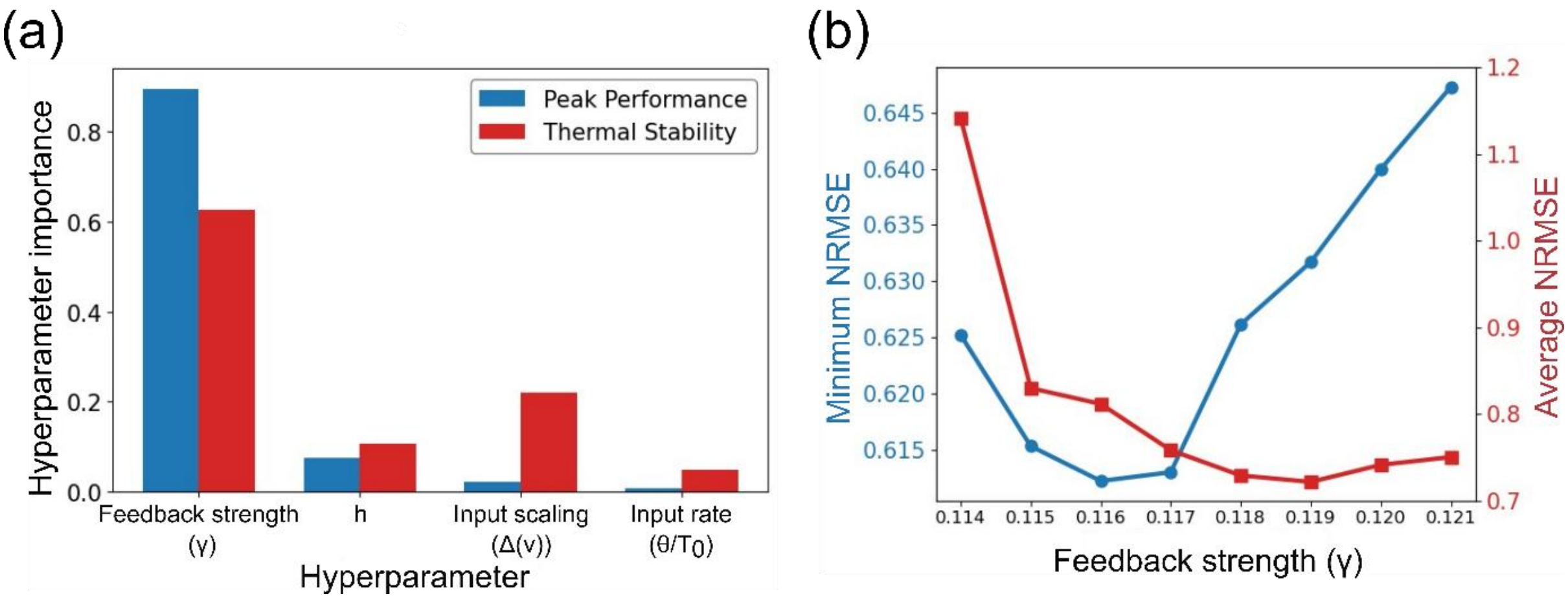

Fig.5. (a) Importances of four hyperparameters for the two objectives. For both objectives feedback strength $\gamma$ exhibits significantly higher importance compared to other hyperparameters. (b) The two objectives as a function of feedback strength. Minimum NMRSE increases with rising $\gamma$, whereas average NMRSE declines.

Fig.5(a) illustrates the importance difference of four hyperparameters (feedback strength $\gamma$, applied field $h$, input scaling $\Delta(v)$ and input rate $\theta/T_0$) for minimum NMRSE and average NMRSE, as quantified through variance decomposition. It is clear that $\gamma$ has the greatest influence on both metrics, with an importance value of approximately 0.9 for minimum NMRSE and 0.6 for average NMRSE, significantly higher than that of other hyperparameters. Consequently, $\gamma$ emerges as the most critical hyperparameter for tuning performance trade-offs.

Fig.5(b) shows the influence $\gamma$ on two objectives. For each run, all parameters of NARMA-10 task and the temperature range explored are identical to those used in Fig. 4. The minimum NMRSE and average NMRSE exhibit different responses to variations in $\gamma$. In the low $\gamma$ region (0.114-0.117), the minimum NMRSE reaches its minimum value of ≈ 0.61 and its value increases significantly as $\gamma$ increases. In this region average NRMSE remains high but decreases as $\gamma$ increases. In high $\gamma$ region (0.118-0.121), average NRMSE is better than that in low $\gamma$ region, with a minimum of ≈ 0.72. However, the minimum NMRSE is increased. These results indicate that $\gamma$ efficiently modulates the trade-off between two objectives, enabling the reservoir to trade-off performance considerations to meet the requirements of diverse application scenarios. For instance, a smaller $\gamma$ value is preferable for the reservoir operating in a temperature-controlled environment such as a laboratory as it provides stronger minimum NMRSE. Conversely, in an outdoor deployment scenario where temperature might vary widely, a larger $\gamma$ could be adopted to prioritize average NMRSE at the cost of marginal minimum NMRSE degradation.

While our study clearly shows that heterogeneous nanodot reservoirs can be optimized via to exhibit excellent average NMRSE of inference performance, two limitations remain: first, we do not have a clear understanding of the quantitative relationship between computational properties and the geometric parameters of the reservoir. Establishing this correlation would deepen our understanding of the reservoir substrate and could provide guidance for further optimization in the heterogeneous structure. Second, the present work has only explores the NARMA-10 task, and does not examine other types of tasks, such as classification and recognition.[32] Different trade-offs may arise across other benchmark tasks, if these impose different requirements on computational properties of the reservoir. Therefore the investigation of task-independent metrics of the reservoir (e.g. memory capacity, kernel rank, and general rank[33,34]) will be important. These may reveal the core computational capabilities of heterogeneous reservoirs rather than their performance on specific tasks.

## Conclusion:

In this paper we have demonstrated that introducing heterogeneity into reservoir computing substrates based on strain-controlled super-paramagnetic nanodot ensembles makes their inference performance robust against ambient temperature fluctuations. In the NARMA-10 benchmark task, a reservoir with the optimised heterogeneity achieved both strong peak task performance at the temperature at which it was trained, and good stability of performance across a temperature range of 5-35°C, in dramatic contrast to homogeneous equivalents. Hyperparameter importance analysis and exploration demonstrated that the reservoir’s feedback strength can be used to tune the trade-off between peak task performance and its average NMRSE, allowing adjustment to different operating conditions. Given these advantages, superparamagnetic nanodot ensembles with heterogeneous geometries represent compelling hardware platforms for reservoir computing in real-world environments.

This work was supported by the Engineering and Physical Sciences Research Council (EPSRC) under Grant Nos. EP/S009647/1, EP/V006339/1, App5646 and App85320. Additionally, this project has received funding from the European Union’s Horizon 2020 research and innovation programme under grant agreement No. 861618 (Spin ENGINE).

## Data availability:

The data that support the findings of this study are openly available in ORDA, https://doi.org/10.15131/shef.data.30940496

# Supplementary material:

# Reproducible reservoir computing with thermally driven superparamagnets: controlling temperature sensitivity

Z. Chen[1], A. Welbourne[1], M. O. A. Ellis[2], D.A. Allwood[1], E. Vasilaki[2], and T. J. Hayward[1]

[1]School of Chemical, Materials and Biological Engineering, University of Sheffield, UK

[2]Department of Computer Science, University of Sheffield, UK

## S1. BEST PREDICTION PERFORMANCE OF UNIFORM NANODOT RESERVOIR

We selected the uniform nanodot ensemble that exhibited the best task performance on the NARMA-10 task as the objective for modelling temperature effects in nanodot reservoirs.

To identify the optimal reservoir, we employed Optuna[1] to perform an automated search for hyperparameters. The parameters and their search ranges are summarized in Table S1. The normalized root mean square error (NRMSE) between the predicted and target NARMA-10 outputs was used as the optimization objective and minimized during optimization. The Tree-structured Parzen Estimator (TPE) sampler was adopted for efficient exploration of the hyperparameter space.

Table S1 Parameters for best uniform reservoir

| Category | Parameter | Unit | Range | Optimal |
|---|---|---|---|---|
| Geometric factors | Diameter ratio ($d_2/d_1$) | - | 0.75 – 1.25 | 0.839 |
| Hyperparameters | Virtual node number ($N_v$) | integer | 20 - 500 | 461 |
| | Feedback strength (γ) | - | 0.05 - 2 | 0.146 |
| | Applied field (h) | - | 0.1 – 0.5 | 0.405 |
| | Mask scaling ($m_0$) | - | 0.001 – 0.01 | 0.007 |
| | Input rate ($\theta/T_0$) | - | 0.2 – 1 | 0.650 |

Fig.S1 presents the parameter search results after 500 trials, where the lowest NRMSE of 0.49 is obtained, in close agreement with our previous results.[2] The hyperparameters of the optimal reservoir are listed in Table S1. The results confirm that the optimized reservoir demonstrates strong predictive capability on benchmark tasks, comparable to other reported reservoirs.[3–5]

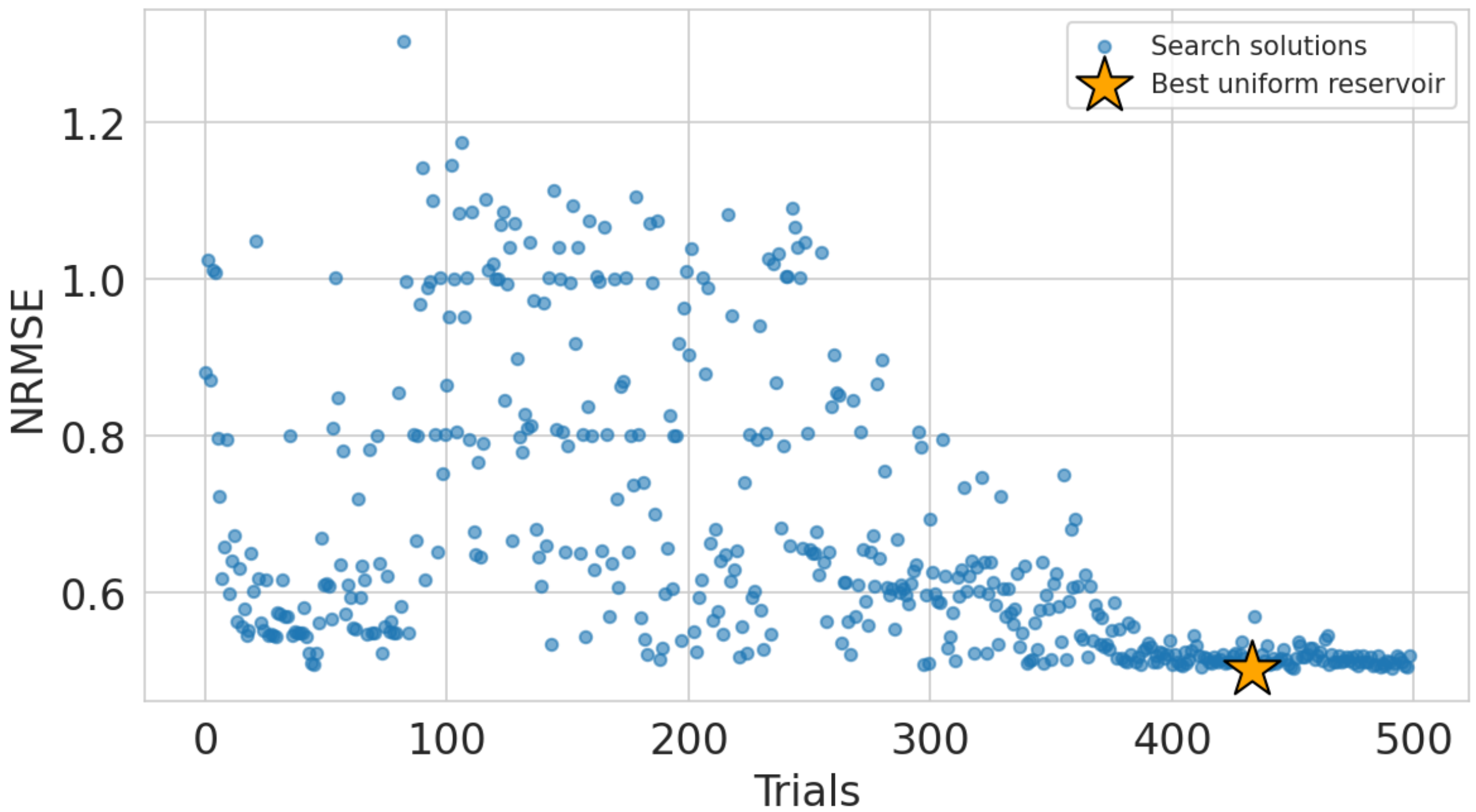

Fig.S1 Hyperparameter search results for obtaining the uniform nanodot ensemble with best task performance on NARMA-10. The yellow star and blue circles present the optimal solution which NRMSE is of 0.49 and other solutions, separately.